\documentclass[onecolumn,showpacs]{revtex4}

\usepackage{graphicx}% Include figure files
\usepackage{dcolumn}% Align table columns on decimal point
\usepackage{bm}% bold math

\input epsf

\begin{document}

\title{\Large A dynamical symmetry of the spherical dust collapse}

\author{\bf Ujjal
Debnath$^1$\footnote{ujjaldebnath@yahoo.com},~Subenoy
Chakraborty$^1$\footnote{subenoyc@yahoo.co.in} and ~Naresh
Dadhich$^2$\footnote{nkd@iucaa.ernet.in}}

\affiliation{ $^1$Department of Mathematics, Jadavpur University,
Calcutta-32, India.\\ $^2$Inter-University Centre for Astronomy
and Astrophysics, Post Bag 4, Pune 411 007, India.}

\date{\today}

\begin{abstract}
By linearly scaling the initial data set (mass and kinetic energy functions)
together with the initial area radius of a collapsing dust sphere, we find a
symmetry of the collapse dynamics. That is, the linear transformation defines
an equivalence class of data sets which lead to the same end result as well as
its evolution all through. In particular, the density and shear remain
invariant initially as well as during the collapse. What the transformation
is exhibiting is an interesting scaling relationship between mass, kinetic
energy and the size of the collapsing dust sphere.
\end{abstract}


\maketitle

In classical general relativity, one of the key outstanding
questions is the end result of a gravitationally collapsing
object, does it end in a black hole (BH) or in a naked
singularity (NS) [1, 2] ? There exists no method to address this
question in general terms, and hence one is left to study various
examples of collapsing systems [3] with a view to gain some
insight. Recently, the most pertinent question, what is the
physical process that causes NS, has also been addressed for
spherical dust as well as for the general Type I matter field
[4,5]. The answer is that it is the shearing force in the
collapsing cloud which is responsible for NS. It has been proven
that shear is necessary for NS while it has to be strong enough
(beyond certain threshold value) to lead to NS.\\

Recently, again for a spherical dust collapse, a linear
transformation has been found which keeps density and shear of
the collapsing sphere invariant for all through the collapse but
for the initial instant [6]. That is the initial density and shear
are not invariant under this transformation. In this letter we
wish to complete the invariance for initial event as well by
bringing in the radius of the sphere as well to scale linearly.
It is now a symmetry of the collapse dynamics and we have an
equivalent set of initial data leading
to the same end result as well as its evolution all through the collapse.\\

The inhomogeneous spherically symmetric dust collapse in $n$
dimensions can be described by the higher dimensional
Tolman-Bondi-Lema\^{\i}tre (TBL) line element given by

\begin{equation}
ds^{2}=-dt^{2}+\frac{R'^{2}}{1+f(r)}dr^{2}+R^{2}d\Omega^{2}_{n-2},
\end{equation}

where the area radius, $R=R(t,r)$ and $f(r)>-1$. The evolution equation for
$R$ is

\begin{equation}
\dot{R}^{2}=\frac{F(r)}{R^{n-3}}+f(r),
\end{equation}

where the arbitrary function $F(r)>0$ is related to the matter
density as

\begin{equation}
\rho(t,r)=\frac{(n-2)F'(r)}{2R^{n-2}R'},
\end{equation}

The measure of anisotropy is given by the shear scalar

\begin{equation}
\sigma=\sqrt{\frac{n-2}{2(n-1)}}\left(\frac{\dot{R}'}{R'}-\frac{\dot{R}}{R}\right)
=\sqrt{\frac{n-2}{8(n-1)}}~\frac{\left[\{R F'-(n-1)R' F
\}+R^{n-3}(R f'-2R' f)\right]}{R^{\frac{n-1}{2}}R'\left(F+f
R^{n-3} \right)^{1/2}}
\end{equation}

Now using the initial condition $R=r$ at $t=t_{i}$ (initial
epoch) the expressions for initial density and shear are

\begin{equation}
\rho_{i}(r)=\rho(t_{i},r)=\frac{n-2}{2}r^{2-n}F'(r)
\end{equation}
and
\begin{equation}
\sigma_{i}=\sigma(t_{i},r)=\sqrt{\frac{n-2}{8(n-1)}}~\frac{\left[\{r
F'-(n-1)F \}+r^{n-3}(r f'-2f)\right]}{r^{\frac{n-1}{2}}\left(F+f
r^{n-3} \right)^{1/2}}
\end{equation}

For the finiteness of the initial density and shear at the
centre, the regular functions $F$ and $f$ must have the following
series expansion

\begin{equation}\begin{array}{cc}
F(r)=F_{0}r^{n-1}+F_{1}r^{n}+F_{2}r^{n+1}+.........~~~~~~~~(F_{0}\ne
0 )
\\\\
f(r)=f_{0}r^{2}+f_{1}r^{3}+f_{2}r^{4}+.........~~~~~~~~~~~~~~~~~(f_{0}\ne
0 )
\end{array}
\end{equation}

For smooth initial density profile we write

\begin{equation}
\rho_{i}(r)=\rho_{0}+\rho_{1}r+\rho_{2}r^{2}+.........
\end{equation}

So using these series expansions in equation (5) we have the relation
among the different coefficients as
$$\rho_{j}=\frac{(n+j-1)(n-2)}{2}F_{j},~~~j=0,1,2,...$$\\

The evolution equation (2) can be integrated to give

\begin{equation}
t-t_{i}=\frac{2}{(n-1)\sqrt{F}}\left[r^{\frac{n-1}{2}}~_{2}F_{1}[\frac{1}{2},a,a+1,-\frac{f
r^{n-3}}{F}]-R^{\frac{n-1}{2}}~_{2}F_{1}[\frac{1}{2},a,a+1,-\frac{f
R^{n-3} }{F}]\right]
\end{equation}

where $_{2}F_{1}$ is the usual hypergeometric function with
$a=\frac{1}{2}+\frac{1}{n-3}$.\\

If $t=t_{s}(r)$ denotes the time of collapse of the $r$-th shell
to the singularity then $R(t_{s}(r),r)=0$ and we have from the
above equation

\begin{equation}
t_{s}(r)-t_{i}=\frac{2}{(n-1)\sqrt{F}}r^{\frac{n-1}{2}}~_{2}F_{1}[\frac{1}{2},a,a+1,-\frac{f
r^{n-3}}{F}]
\end{equation}

The formation of trapped surface is signaled by $R^{n-3}=F(r)$ at
$t=t_{ah}(r)$ and so we write,

\begin{equation}
t_{ah}(r)-t_{i}=\frac{2r^{\frac{n-1}{2}}}{(n-1)\sqrt{F}}~_{2}F_{1}[\frac{1}{2},a,a+1,-\frac{f
r^{n-3}}{F}]-\frac{2F^{\frac{1}{n-3}}}{n-1}~_{2}F_{1}[\frac{1}{2},a,a+1,-f]
\end{equation}

Using the series expansion (7) if we proceed to the limit as
$r\rightarrow 0$ in equation (10) then we have the limit of
central singularity as

\begin{equation}
t_{0}=t_{i}+\frac{2}{(n-1)\sqrt{F_{0}}}~_{2}F_{1}[\frac{1}{2},a,a+1,-\frac{f_{0}
}{F_{0}}]
\end{equation}

Hence the time difference between the formation of apparent
horizon and central singularity is given by (upto leading order in
$r$)

\begin{eqnarray*}
t_{ah}-t_{0}=-\frac{2}{n-1}F_{0}^{\frac{1}{n-3}}r^{\frac{n-1}{n-3}}-
\frac{r}{(n-1)\sqrt{F}}\left[\frac{F_{1}}{F_{0}}~_{2}F_{1}[\frac{1}{2},a,a+1,-\frac{f_{0}
}{F_{0}}]\right.
\end{eqnarray*}

\begin{equation}
\left.+\frac{(n-1)}{(3n-7)}\frac{f_{0}}{F_{0}}\left(\frac{f_{1}}{f_{0}}-\frac{F_{1}}{F_{0}}\right)
~_{2}F_{1}[\frac{3}{2},a+1,a+2,-\frac{f_{0} }{F_{0}}]\right]
\end{equation}

This reduces in $4$ and $5$ dimensions respectively to

\begin{eqnarray*}
t_{ah}-t_{0}=\frac{1}{10}\left[\frac{5F_{1}}{f_{0}^{3/2}}~sinh^{-1}\sqrt{\frac{f_{0}}{F_{0}}}
-\frac{5F_{1}\sqrt{1+\frac{f_{0}}{F_{0}}}}{f_{0}\sqrt{F_{0}}}\right.
\hspace{1in}
\end{eqnarray*}

\begin{equation}
\left.+\frac{2(f_{0}F_{1}-f_{1}F_{0})}{F_{0}^{5/2}}~
_{2}F_{1}[\frac{3}{2},\frac{5}{2},\frac{7}{2},-\frac{f_{0}
}{F_{0}} \right]~r+O(r^{2})
\end{equation}

\begin{equation}
t_{ah}-t_{0}=\left[\frac{(2f_{1}F_{0}-f_{0}F_{1})}{2f_{0}^{2}\sqrt{F_{0}}}+
\frac{(f_{0}F_{1}-f_{0}f_{1}-2f_{1}F_{0})}{2f_{0}^{2}\sqrt{F_{0}}\sqrt{1+\frac{f_{0}}{F_{0}}}}
\right]~r+O(r^{2})
\end{equation}

Obviously it is NS when $t_{ah}>t_{0}$ and BH otherwise. It is
the data set $\{F,f\}$ that determines the ultimate end result in
terms of BH or NS for the collapsing sphere.\\

Mena etal [6] have proposed the following linear transformation on
the data set $I=\{F,f\}$,
\begin{equation}
\{F,f\}\rightarrow \left\{a^{\frac{n-1}{2}}F,af \right\}
\end{equation}
($a>0$  a constant) under which both $\rho,\sigma$ remain
invariant but not initially, i.e $\rho_{i},\sigma_{i}$ are not
invariant. They have however noted that it is this discontinuous
behaviour which is responsible for having NS for initially
vanishing shear or density contrast. It is this discontinuity
which compensates for absence of initial density contrast or
shear. It is the reflection of the fact that it is always
possible to generate an NS for any arbitrary density profile by
suitably choosing the energy function $f$ [7]. It is however
interesting that this transformation also picks up this general aspect. \\

Our concern here is to seek a symmetry of the collapse dynamics -
a true invariance. It stands to reason that when we linearly
scale mass and energy functions, we should also scale the radius
$R$ of the sphere to keep everything in balance. The above
transformation should thus be supplemented by
\begin{equation}
R \rightarrow \sqrt{a} ~R.
\end{equation}
Now this is the symmetry of the complete collapse dynamics. It
can be easily verified that the density and shear both initially
as well as otherwise do remain unaltered under the
transformation. Also the time difference $(t_{ah} - t_{0})$
remains invariant indicating the complete symmetry of the collapse
process. For marginally bound collapse we can set $f=0$ in the
transformation without disturbing the invariance.\\

We have thus completed the linear transformation proposed by Mena
etal [6] to make it as a symmetry of the collapse dynamics. What
it indicates is that if we scale the radius $R$ by $\alpha$, then
mass function $F$ will naturally scale as $\alpha^{n-1}$ and the
energy $f$ as $\alpha^2$ (dimensionally $F \rightarrow
R^{(n-3)}F$). The transformation is therefore physically quite
transparent but the remarkable fact is that the overall dynamics
also respects it all through. This is the dynamical symmetry of
spherical collapse. It is though established for the dust collapse
but it should be true in general for other matter
fields as well. \\\\

{\bf Acknowledgement:}\\

One of the authors (U.D) is thankful to CSIR (Govt. of India) for
awarding a Senior Research Fellowship and both SC and UD
acknowledge the hospitality of IUCAA.\\


\begin{thebibliography}{99}
\bibitem{sw} S.W. Hawking and G.F.R. Ellis, {\it The large scale structure
of space-time} (Cambridge. Univ. Press,
Cambridge, England, 1973).\\
\bibitem{pen}  R. Penrose, {\it Riv. Nuovo Cimento} {\bf 1} 252 (1969);
in {\it General Relativity}, an Einstein
Centenary Volume, edited by S.W. Hawking and W. Israel (Camb. Univ. Press, Cambridge, 1979).\\
\bibitem{psj}  For recent reviews, see, e.g. P.S. Joshi, {\it Pramana}
{\bf 55} 529 (2000); C. Gundlach, {\it Living Rev. Rel.} {\bf 2}
4 (1999).\\
\bibitem{jdm} P.S. Joshi, N. Dadhich and R. Maartens, {\it Phys. Rev. D} {\bf 65} 101501({\it R})(2002).\\
\bibitem{jgm} P. S. Joshi, R. Goswami and N. Dadhich, {\it The critical role
of shear in gravitational collapse - II}, gr-qc/0308012, { Phys. Rev. D} in
press.\\
\bibitem{mena} F. C. Mena, B. C. Nolan and R. Tavakol, {\it gr-qc}/0405041.\\
\bibitem{dwi} I.H. dwivedi and P. S. Joshi, {\it Class. Quant. Grav.} {\bf 14}
1223 (1997).\\
\end{thebibliography}
\end{document}